# A Hybrid Model of Rubber Elasticity in Simple Extension

Arkady I. Leonov

*Department of Polymer Engineering, The University of Akron,* Akron, OH 44325-0301

E-mail address: leonov@uakron.edu (A.I. Leonov)

**Abstract**

A thermodynamically related model is developed for describing elastic rubber-like behavior of amorphous and crystallizing polymers and demonstrated on example of simple extension. Both the "entropic" and "energetic" motions of polymer chains that contribute in macroscopic elastic deformation are taken into account. The model displays a continuous transition from entropy to energetic elasticity, without common singularity caused by finite extensibility of polymer chains. A multi-scale molecular approach, based on recent literature concepts has been employed for evaluations of continuum parameters. For crystallizing polymers, a simple model is developed for the stress-induced crystallization, which describes the stress reinforcement caused by the formation of long, needle-like polymer crystals.

## 1. Introduction

Amorphous and crystallizing polymers demonstrate different types of mechanical and optical behavior above $T_g$. The amorphous polymers display the rubber-like behavior in both the cross-linked and not cross-linked states, the latter just above $T_g$. In case of cross-linked rubbers, the importance of energetic component of deformation has been discussed long ago (e.g. see [1]), but to the best author knowledge, the energetic effects have never been involved in rubber theories. In case of not cross-linked polymers, the energetic component(s) of deformation, along with common entropy component, could also significantly contribute in reversible part of stress tensor in rubbery region due to the high level viscosities in the low temperature region above $T_g$. Thus a "hybrid", energetic/entropy approach could facilitate understanding of specifics of rubbery behavior. A hybrid continuum constitutive model [2,3] of relaxation type with additive energetic and entropy stresses has been elaborated for describing rheological behavior of amorphous polymers near the glass transition.

A new molecular concept, based on recent experimental results (e.g. see [4,5]) could in principle help in microscopic understanding of macroscopic processes that might be involved in a possible hybrid behavior. The papers [4,5] among others, revealed that in the well-known bid-spring model [6,7] of polymer chains, a set of statistically independent bids should not be treated as commonly viewed simplifying mathematic abstraction, but some real parts of macromolecules ('sub-chains", or "dynamic segments"). Thus the bids should have certain masses with a certain average number of monomer units $n_d$, specific for a given polymer. The concept of Kuhn segments [1], as statistically independent units introduced earlier in rubber elasticity, has been recently criticized [8] as inconsistent with many experimental data. The earlier results of semi-empirical statistical model and analyses of flow data for polymer melts [9] also showed that the concept of Kuhn segment could not explain well-known fact of independence of activation energy of the length of polymer chain. Therefore the (flow or) *dynamic segment* was defined in [9] as a part of macromolecule that participates in overcoming energetic (rotational) barriers with highly correlated, cooperative motions of monomer units, when keeping the connectivity of macromolecular chains. A similar concept with different scaling calculations has also been recently proposed [10] to explain high frequency dynamic data for dilute polymer solutions. The ideas displayed in papers [9,10] can be interpreted as a rough attempt to understand the cooperative motions of monomer units inside the dynamic segments as the source of "energetic elasticity". This type of elasticity has been mentioned many times in the literature (e.g. see [1-3]. Several more fundamental theoretical and computational approaches have also been developed to understand the kinetics of conformational transitions of parts of macromolecules on microscopic level (e.g. see [11-13]). However, the most important effect of cooperativeness, more difficult for theoretical treatment, has not been understood.

In simple extension of crystallizing polymers, the stress-induced crystallization could happen at high stretching ratios. The needle-like (NL) polymer crystals that emerge in this type of crystallization have been observed long ago in higher extension of natural and synthetic crystallizing cross-linked rubbers [1](ch.1). These crystals are quite different from the common folding crystals emerged under cooling in the stress free crystallization. Flory [14] developed a statistical model for the stress-induced crystallization based on the classical entropy (Gaussian) statistical calculations that neglected the non-Gaussian effects of finite extensibility of polymer chains, and internal energy contribution. The dependence of extension force on stretching ratio in stress-induced crystallization [14] displayed decreasing stress in crystalline region as compared to that for amorphous case. However, the experimental data [15,16] clearly demonstrated exactly opposite trend of increasing stress in the crystalline region and were explained as reinforcement of rubber by the emerged rigid crystals. Denying this explanation, Flory [14] argued that this discrepancy is due to the irreversible relaxation effects caused by continuous deformations. Later, Flory's theory has been modified with introducing two new aspects: (i) nonalignment of the crystalline chains along the tensile axis [17], and (ii) combination of NL and common folding-like crystallization in cooling [18]. It should be mentioned that the modification (ii) could contradict the experimental data where the opposite effects, unfolding the folded polymer crystals under stretching, were reported (e.g. see [19,20] and references there).

This paper is organized as follows. We first develop a general equilibrium, continuum approach for simple extension to obtain the hybrid stress-strain constitutive relation, and then specify thermodynamic functions responsible for entropic and energetic contributions. Then using scaling approach based on the multi-scale molecular model, we evaluate macroscopic parameters in the continuum model. In the final part of the paper we extend the previous results on crystallizing polymers. Here we first develop a local thermodynamic description of formation of NL crystals and evaluate their parameters. We then develop a mostly phenomenological continuum model of stress-induced crystallization. Only 1D simple extensional approach is presented in the paper. The full 3D tensor formulation of the hybrid rubber elasticity will be published elsewhere.

## 2. Qualitative Description of Structure

Underlying the scaling approach in this paper is a *multi-scale* molecular model, which is qualitatively described as follows. On the *large scale*, macromolecules are viewed as consisting of $N_d$ freely jointed and toughly deformable "dynamic segments" (or "bids") consisting of $n_d$ monomer units so that $N = N_d n_d$. Here $N$ is the degree of polymerization. The evaluation of $n_d$ can be found in papers [8-10] and references there. An important geometrical parameter of dynamic segments is their initial, non-deformed length $l_d$, which highly increases with the increase in the chain rigidity factor [9]. Unlike the definition of the Kuhn segment, in any case $l_d N_d < L$ (or even $\ll L$), where $L$ is the contour chain length. Thus at moderate stretching ratios, the large-scale motion responsible for the entropy elasticity can be viewed as a conformational motion of macromolecule consisting of freely joined dynamic segments (bids) of length $l_d$.

On the *small scale*, the motions of monomer units inside the dynamic segments are due to stretching of small parts of macromolecules. Therefore the segments can be schematically viewed, as toughly stretching bids, needed higher energies for their deforming. When the entropy elasticity is exhausted (and even before that), the dynamic segments, being almost oriented in the stretching direction, start intensively deforming due to the motions of monomer units collectively overcoming rotational barriers (while preserving chain connectivity).

When the monomer units in a dynamic segment are maximally oriented along the stretched chains, another, *fine-scale,* motion of monomer units is still possible. This motion, caused by distortion of valence angles between adjacent monomer units, is responsible for the solid like infinitesimal elasticity of the same type as in low molecular weight crystals.

Both the small- and fine-scale deformations are two different types of *energetically related* elasticity. Nevertheless, we will use in the following the term "energetic elasticity" only for the less tough energetic deformations related to collective motions of monomer units overcoming the *rotational* barriers. The other, extremely tough energetic elasticity caused by distortion of valence angles between adjacent monomer units will be called as the crystal-like (CL) elasticity.

On the continuum level, there are three macroscopic Hookean moduli, entropic $G_s$, energetic $G_e$, and CL one $G_c$, that characterized respectively these three types of elasticity, such that

$$G_s \ll G_e \ll G_c. \tag{1}$$

In summary, the present model views the polymer chains as consisted of *free jointed, deformable dynamic segments.* The large- small- and scale motions of monomer units produce on macroscopic level the entropic and energetic components of deformations, which are not independent. The behavior of the deformable bids in two-scale (entropic/energetic) bid-spring model is schematically illustrated in Figure 1.

The above multi-scale model can also be mechanistically pictured as three consecutively connected springs, the first one ("entropic") being soft, another ("energetic") tough and the third one ("CL") extremely tough. If this hybrid spring is extended by a relatively small force $F$, spring's full displacement $\Delta l = \Delta l_e + \Delta l_s + \Delta l_c$ is the sum of entropy $\Delta l_s = F/\kappa_s$, energetic $\Delta l_e = F/\kappa_e$, and CL $\Delta l_c = F/\kappa_c$, components, where $\kappa_s, \kappa_e$ and $\kappa_c$ are respectively the entropy, energetic and CL spring constants. The hybrid's force-displacement relation $F = \kappa \Delta l$ with the hybrid spring constant $\kappa = (\kappa_s^{-1} + \kappa_e^{-1} + \kappa_c^{-1})^{-1}$ clearly shows that under condition $\kappa_s \ll \kappa_e \ll \kappa_c$, the entropy spring overwhelmingly contributes in the total displacement of the spring.

**2. General Thermodynamic Model**

We neglect here for simplicity the volume deformations, i.e. consider $\rho = \rho_0 = const$. We assume that in the thermodynamic relation $\Delta f = \Delta e - T\Delta s$, where $f$ is the Helmholtz free energy (density), the changes in internal energy $\Delta e$ and entropy $\Delta s$ are represented as:

$$\rho_0 \Delta e = 1/2 G_e(T)\psi_e(\lambda_e) + 1/2 G_c(T)\psi_c(\lambda_c), \quad -\rho_0 \Delta sT = 1/2 G_s(T)\psi_s(\lambda_s). \tag{2}$$

Here $\lambda_s$, $\lambda_e$ and $\lambda_c$ are, respectively, the entropic, energetic and CL elastic stretching ratios. Equation (2) means:

$$\rho_0 f(\lambda_e, \lambda_s, T) = 1/2 G_s(T)\psi_s(\lambda_s) + 1/2 G_e(T)\psi_e(\lambda_e) + 1/2 G_c(T)\psi_c(\lambda_c). \qquad (3)$$

The specific dependencies (2) and (3) of internal energy and entropy on their own, stretching components is the main assumption of the hybrid approach. It is physically meaningful only if the inequalities (1) are valid.

The relation between the entropic $\lambda_s$, energetic $\lambda_e$, CL $\lambda_c$ and total $\lambda$ stretching ratios, evident from the picturesque of the multi-scale model, is postulated as:

$$\lambda = \lambda_s \cdot \lambda_e \cdot \lambda_c. \qquad (4)$$

In the regions of dominancy of either entropic or energetic elasticities, the respective true stresses are represented as:

$$\sigma_s(\lambda_s,T) = 1/2 G_s \lambda_s d\psi_s/d\lambda_s, \quad \sigma_e(\lambda_e,T) = 1/2 G_e \lambda_e d\psi_e/d\lambda_e, \quad \sigma_c(\lambda_c,T) = 1/2 G_c \lambda_c d\psi_c/d\lambda_c \qquad (5)$$

According to the physical sense of the hybrid modeling, the entropic and energetic stresses are not additive but related as:

$$\sigma_s(\lambda_s,T) = \sigma_e(\lambda_e,T) = \sigma_c(\lambda_c,T) = \sigma(\lambda,T). \qquad (6)$$

Here $\sigma(\lambda,T)$ is the total stress.

It is easy to show that the relations (3)-(6) are compatible with the common definition of the total stress-stretch relation,

$$\sigma(\lambda,T) = \rho_0 \lambda \partial f/\partial \lambda \big|_T. \qquad (7)$$

Indeed, calculating the right-hand side (7) with account of (3) yields:

$$\sigma = \rho_0 \lambda \partial f/\partial \lambda \big|_T = 1/2 \lambda_s \lambda_e \lambda_c \sum_{i=s,e,c} G_i (d\lambda_i/d\lambda) d\psi_i/d\lambda_i$$
$$= \sigma_s \lambda_e \lambda_c \partial \lambda_s/\partial \lambda + \sigma_e \lambda_s \lambda_c d\lambda_e/d\lambda + \sigma_c \lambda_e \lambda_s \partial \lambda_c/\partial \lambda.$$
$$= \sigma(\lambda_e \lambda_c d\lambda_s/d\lambda + \lambda_s \lambda_c d\lambda_e/d\lambda + \lambda_s \lambda_e d\lambda_c/d\lambda) \equiv \sigma$$

Here we used the identity,

$$\lambda_e \lambda_c d\lambda_s/d\lambda + \lambda_s \lambda_c d\lambda_e/d\lambda + \lambda_s \lambda_e d\lambda_c/d\lambda \equiv 1,$$

obtained by differentiating (4) with respect to $\lambda$. This derivation clearly shows that in elastic case, the relations (4) and (7) are inconsistent with the assumption of stress additivity, $\sigma = \sigma_e + \sigma_s$, employed in viscoelastic case in [2,3].

The constitutive equation, $\sigma = \sigma(\lambda, T)$, can now be readily established using the functions $\sigma_e(\lambda_e, T)$ and $\sigma_s(\lambda_s, T)$ defined in (5). Assuming that the hybrid model is thermodynamically stable results in the fact that the three stress functions $\sigma_s(\lambda_s, T)$, $\sigma_e(\lambda_e, T)$ and $\sigma_c(\lambda_c, T)$ are monotonically increasing. Then using (6), one can introduce the three inverse functions, $\lambda_s = \sigma_s^{-1}(\sigma, T)$, $\lambda_e = \sigma_e^{-1}(\sigma, T)$ and $\lambda_c = \sigma_c^{-1}(\sigma, T)$, and obtain with the aid (4) the inverse hybrid CE in the general form:

$$\lambda(\sigma, T) = \sigma_s^{-1}(\sigma, T) \cdot \sigma_e^{-1}(\sigma, T) \cdot \sigma_c^{-1}(\sigma, T). \tag{8}$$

In order to specify relation (8), some physical models for entropic, energetic and CL elasticity should be introduced.

## 3. Specific Thermodynamic Model and Hybrid CE's

For deriving a hybrid CE, we first need to model the three types of elasticity, entropic, energetic and crystal-like ones.

*3.1. Entropy elasticity*

When the Gaussian statistics is valid, there are the familiar expressions for the classic entropic elasticity:

$$-\rho_0 (s - s_0) T \equiv W_s(\lambda_s, T) = 1/2 G_s(T)(\lambda_s^2 + 2/\lambda_s - 3), \quad \sigma = G_s(T)(\lambda_s^2 - 1/\lambda_s) \quad . \tag{9}$$

When stretching ratio is high enough and non-Gaussian effects are important, fractioning of macromolecular chains in the dynamic segments, suggested by the multi-scale model, plays a pivotal role. That is because as compared to the Kuhn segment statistics, the onset of non-Gaussian behavior for dynamic segments begins much earlier for chains consisting of dynamic segments that are as a rule much larger than the Kuhn segment [9]. In the following, we simply use the semi-empirical Warner-Gent potential [14,15], which describes the finite extensibility of polymer chains in terms of Finger strain tensor. In simple extension, the elastic potential and related stress are presented as:

$$-\rho \Delta s T \equiv W_s(\lambda_s, T) = -1/2 G_s(T)(I^* - 3) \ln[1 - (I_{1s} - 3)/(I^* - 3)] \quad (I_{1s} = \lambda_s^2 + 2/\lambda_s). \tag{10a}$$

$$\sigma = \sigma_s(\lambda_s, T) = G_s(T)(\lambda_s^2 - 1/\lambda_s)[1 - (I_{1s} - 3)/(I^* - 3)]^{-1}. \tag{10b}$$

Here, $I^* = I_{1s}(\lambda_*)$, and $\lambda_*$ is the strain related to complete aligning the dynamic segments but *without taking into account their internal stretching*. Keeping in mind that the analysis of entropy elasticity in our hybrid model is based on the fractioning of polymer chains in large enough dynamic segments, one can conclude that the value $I^*$ in (10) might be considerably lower than that when using the Kuhn segment. Note that involving in this model a possible dependence of $\lambda_*$ on $\lambda_e$, which contradicts the basic assumption (2), makes the analysis needlessly more complicated.

For $\lambda_s \gg 1$ (and always for $\lambda_* \gg 1$), the stress in (10b) is expressed in the simplified form:

$$\sigma = \sigma_s(\lambda_s, T) \approx G_s(T) \frac{\lambda_s^2}{1 - (\lambda_s/\lambda_*)^2}. \quad (\lambda_s \gg 1) \tag{10c}$$

3. 2. *Energetic elasticity*

To characterize this type of elasticity we employ below the simplest approach of the finite solid elasticity made in terms of Cauchy-Green strain tensor. This approach is valid even for the relatively strong energetic case when $n_d \gg 1$. In simple extension, it is presented as:

$$W_e(\lambda_e, T) = 1/2 G_e(T) \psi_e(\lambda_e) = 1/2 G_e(T)(2\lambda_e + \lambda_e^{-2} - 3) \tag{11a}$$

$$\sigma = G_e(T)(\lambda_e - \lambda_e^{-2}) \tag{11b}$$

In respective cases of *weak and strong energetic elasticity,* relation (11) for stress takes the asymptotic forms:

$$\sigma \approx 3G_e(\lambda_e - 1) \quad (\lambda_e - 1 \ll 1), \tag{11c}$$

$$\sigma \approx G_e \lambda_e \quad (\lambda_e \gg 1). \tag{11d}$$

3.3. *CL elasticity*

This is infinitesimal elasticity, where

$$W_c(\lambda_c, T) = 1/2 G_c(T) \psi_c(\lambda_c) = 3/2 G_c(T) \varepsilon_c^2 \tag{12a}$$

$$\sigma = 3G_c(T)\varepsilon_c, \quad \lambda_c = 1 + \varepsilon_c \quad (0 < \varepsilon_c \ll 1). \tag{12b}$$

Note that the temperature dependencies $G_s(T)$ and $G_e(T)$ (or $G_c(T)$) are qualitatively different. Unlike slightly increasing function $G_s(T)$, the energetic moduli $G_e(T)$ and $G_c(T)$ are slightly decreasing function of $T$, having a maximum about $T_g$ and being independent of $T$ at higher temperatures in the rubbery region.

3.4. *Hybrid CE's.*

Formula (8) along with the expressions for stress in (10) and (11) constitutes the hybrid constitutive relation $\sigma(\lambda)$, that could rarely be explicitly expressed. Two cases will be considered to illustrate this constitutive behavior under the conditions (1).

We first consider the transition from entropic to the energetic elasticity assuming that the contribution of the CL elasticity in stress is negligible (i.e. that "crystal-like" springs are rigid). To illustrate the features of this transition we consider the large entropic strain approximations (11d). Then CE (8) for hybrid elasticity takes the forms:

$$\hat{\lambda} \approx \left(1 + \frac{\hat{\sigma}}{3\alpha}\right)\sqrt{\frac{\sigma}{1+\hat{\sigma}}} \quad (\hat{\sigma} \ll \alpha); \quad \hat{\lambda} \approx \frac{\hat{\sigma}}{\alpha}\sqrt{\frac{\sigma}{1+\hat{\sigma}}} \quad (\hat{\sigma} \gg \alpha) \qquad (13a,b)$$

$$(\hat{\lambda} = \lambda/\lambda_*, \quad \hat{\sigma} = \sigma/(G_s\lambda_*^2), \quad \alpha = G_e/(G_s\lambda_*^2))$$

Here parameter $\alpha$ is the measure of relative contributions of entropic and energetic elasticity effects in the total hybrid elasticity. When $\hat{\sigma} \ll \alpha$, due to (13a) $\lambda_e \approx 1$, i.e. the energetic contribution in the hybrid elasticity is negligible. Therefore this case is formally described by (10c) with $\lambda_s \to \lambda$. When $\hat{\sigma} \gg \alpha$, formula (13b) well describes the hybrid elasticity with large stretches for arbitrary value of parameter $\alpha$. At the crossover point $\hat{\lambda}_\alpha$ between two asymptotic behaviors in (13), the stress and strain characteristics are:

$$\hat{\lambda} = \lambda_\alpha: \quad \lambda_\alpha = \lambda_* \frac{3}{2}\sqrt{\frac{3\alpha/2}{1+3\alpha/2}}, \quad \sigma_\alpha = \frac{3}{2}G_e, \quad \lambda_s^{(\alpha)} = \lambda_*\sqrt{\frac{3\alpha/2}{1+3\alpha/2}}, \quad \lambda_e^{(\alpha)} = \frac{3}{2}. \qquad (14)$$

Consider now the dependence $\sigma(\lambda;\alpha)$ of the hybrid CE on parameter $\alpha$. When $\alpha \gg 1$, the dynamic segments are almost non-deformable up to very high stresses. Thus in this case the entropy elasticity dominates over large deformation region. When $\alpha \leq 1$, the dynamic segments are significantly deformable even at moderate strains. The behavior of dependence $\hat{\lambda}(\hat{\sigma})$ for various values of $\alpha$ is sketched in Figure 2.

It is clearly seen from the above analysis in general, and relation (13) in particular, that the present hybrid model describes a smooth transition from the pure entropy elasticity to the pure energetic one, with no singular behavior common for pure entropy approach with finite extensibility of polymer chains.

The situation when the stretching of chains in dynamic segments is close to their limit is still well described by the relation (13b). Using (11d) where $\hat{\sigma} \gg 1$, this relation can be equivalently represented in the form:

$$\lambda \approx \lambda_* \cdot \lambda_e, \text{ or } \sigma \approx G_e \lambda / \lambda_* \quad (\sigma \gg G_s \lambda_*^2). \tag{15a}$$

Formula (15a) represents the intermediate asymptotic case when the entropy elasticity is already exhausted $(\sigma \gg G_s \lambda_*^2)$ but the stretching effect of the CL elasticity is still very small. If the latter effect (however it is small) is not neglected, the asymptotic formula of hybrid elasticity is of the form:

$$\lambda \approx \lambda_* \cdot \lambda_e \cdot (1 + \varepsilon_c). \tag{15b}$$

Consider now the transition from the energetic to the CL elasticity. Near the transition from energetic to CL elasticity, formula (11d) is generally invalid (see discussion in the next Section). Nevertheless, we can still approximately describe this transition using (15b) under assumption that near this transition the energetic elasticity is almost exhausted. It means that in this limit situation, the chains in the dynamic segment are almost completely extended, with $\lambda_e \approx \lambda_{em}$. Here $\lambda_{em}$ is a maximal stretch of dynamic segment whose value is evaluated in the next Section. As soon as the equality $\lambda_e \approx \lambda_{em}$ is achieved, the CL elasticity, however it is small, cannot be ignored. Then using (15b) and (12b) the strain-stress relation in the region of the CL elasticity, is given by:

$$\lambda \approx \lambda_* \cdot \lambda_{em}(1+\varepsilon_c), \text{ or } \sigma = 3G_c\left(\frac{\lambda}{\lambda_* \cdot \lambda_{em}} - 1\right). \quad (\lambda > \lambda_{em}) \tag{15c}$$

The relation (15c) approximately describes the continuous transition from the energetic to CL elasticity with a kink at $\lambda_e \approx \lambda_{em}$.

## 4. Scaling Evaluation of Continuum Parameters

Four parameters have been employed in the above continuum modeling for general description of equilibrium mechanics of rubbery polymers: (i) the modulus of entropy elasticity $G_s$, (ii) the stretching ratio $\lambda_*$ for finitely (non-Gaussian) extendable polymer coils, fractioned in the non-deformed dynamic segments of length $l_d$ (that should itself be evaluated), (iii) the energetic modulus $G_e$, and (iv) the modulus $G_c$ of CL elasticity. These parameters are evaluated below, using simple scaling arguments that follow from the above multi-scale modeling.

(*). *Evaluation of the equilibrium (non-stretched) length $l_d$ of dynamic segments* needs complicated theoretical/computational studies of highly cooperative motions of monomer units, collectively overcoming rotational barriers. Those motions happen in an effective field caused by the attractive/repulsive forces between monomer units. If the attractive forces dominate over the repulsive ones, dynamic segment have on average a spherical (globular) shape, where $l_d \sim l\, n_d^{1/3}$ and $l$ being the effective length of monomer unit. In case when repulsive forces are essential, the value of $l_d$ can be roughly evaluated as the gyration diameter $d_d \approx l_d$ of the dynamic segment consisting of the $n_d$ monomer units, i.e. $l_d \sim l\, \sqrt{n_d}$. Thus generally,

$$l_d \sim l\, n_d^p. \quad (1/3 \le p \le 1/2) \tag{16}$$

(i). *Evaluation of entropy elasticity modulus $G_s$*, made for dynamic segments in the same way as for Kuhn segments, yields the expression well-known from the rubber elasticity:

$$G_s(T) = \rho RT / M_c. \tag{17}$$

(ii). *Evaluation of stretching ratio $\lambda_*$ for finitely (non-Gaussian) extensible polymer coils consisted of the dynamic segments.* With known length $l_d$ of dynamic segment, the gyration diameter $D_d$ of the polymer coil consisting of the $N_d$ dynamic segments, and the *maximum conformational length* $L_{conf}$, i.e. the length of completely aligned chain consisted of not stretched dynamic segments, are easily evaluated as:

$$D_d \sim l_d \sqrt{N_d}, \quad L_{conf} \approx l_d N_d. \tag{17}$$

Then using (17) the value $\lambda_*$ is evaluated, as:

$$\lambda_* \approx L_{conf} / D_d \sim \sqrt{N_d} \ . \tag{18}$$

(iii) *Evaluation of energetic modulus* $G_e$. The highly stretched amorphous polymers with dominant energetic behavior could be viewed as being consisting of almost extended set of dynamic segments. In the asymptotic situation when the dynamic segment is fully extended to the maximum limit length $l_m = l\, n_d$, the energetic free energy $e$ spent for deformation, is fully compensated by the energy the monomer units spent for overcoming the rotational barrier $\Delta E_r$, so that

$$1/2 G_e \psi_e (\lambda_{em}) \approx \nu \Delta E_r . \tag{19}$$

Here $\nu$ is the number of monomer segments in unit volume and $\lambda_{em}$ is the ultimate energetic stretching ratio defined as:

$$\lambda_{em} \approx l_m / l_d = l\, n_d / l_d \approx n_d^{1-s} . \quad (1/3 \le s \le 1/2) \tag{20}$$

. Using (19), the energetic modulus $G_e$ is then evaluated by the following equivalent relations:

$$G_e \approx \frac{2\nu \Delta E_r}{\psi_e(\lambda_{em})} = \frac{2\rho \Delta E_r}{m\psi_e(\lambda_{em})} \approx \frac{2\Delta E_r}{l^3 \psi_e(\lambda_{em})} = \frac{2\Delta E_a}{l^3 n_d \psi_e(\lambda_{em})} . \tag{21}$$

Here $\rho$ is the density and $m$ is the molecular weight of monomer unit. As seen, the approximate "mechanistic" relation $m \approx \rho l^3$ has been used in (21).

When $\lambda_{em} \gg 1$, one can see that due to (11) $\psi_e(\lambda_{em}) \approx 2\lambda_{em}$. In this case the formulae in (21) take the particular form:

$$G_e \approx \frac{\rho \Delta E_r}{m\lambda_{em}} \approx \frac{\Delta E_r}{l^3 \lambda_{em}} \approx \frac{\Delta E_a}{l^3 n_d^{2-s}} . \tag{22}$$

We remind that in (21) and (22), $\Delta E_a$ ($\approx n_d \Delta E_r$) is the (Arrhenius) activation energy of viscous flow.

(iv) *Evaluation of CL modulus* $G_c$. Near the transition from energetic to CL elasticity the energetic and CL stretching parameters cannot be considered as independent. Nevertheless, using the simplified molecular modeling in [12], we still can obtain a rough evaluating of $G_c$. A phantom chain model was used in [12] where three main motions of bonds (or monomer units) were considered: (a) rotational motion (already considered as contributed in the energetic elasticity), (b) motion due to variation in the bond length described by a simple harmonic potential, and (c) motion due to the distortion of valence angle $\theta_0$. Since the energies of bond stretching are very high as compared to the energy of $\theta$ distortion, we consider only the motions of (c) type, described for a single bond by the following potential:

$$U(\theta) = \frac{1}{2} u_\theta (\cos\theta - \cos\theta_0)^2 \approx \frac{u_\theta \sin^2\theta_0}{2}(\delta\theta)^2 . \qquad (23)$$

Here $\theta_0$ is the valence angle between two neighboring bond vectors and $\delta\theta \equiv \theta - \theta_0$ is the valence angle distortion for the bond. We considered the potential (23) near the minimum, $\theta = \theta_0$ for the fully stretched chain, when the variations in the expressions (23) for different bonds are negligible. The force applied along the fully stretched chain causes the longitudinal displacement of chain, calculated as the sum of equal displacements due to the distortion of valence angles in each bond. The axial displacement for each bond is calculated as

$$l_c - l_0 \equiv b[\cos(\theta/2) - \cos(\theta_0/2)] \approx 1/2 b \cdot \delta\theta \cdot \sin(\theta_0/2).$$

For the zigzag configuration of chain, the stretching ratio is:

$$\varepsilon_c = (l_c - l_0)/l_0 = 1/2(\delta\theta)\tan(\theta_0/2).$$

Here $l_c$ and $l_0$ are the projections of the bond on zigzag axis with disturbed and undisturbed values of the valence angle, respectively. For the helix configuration, the result is almost the same. Expressing $\delta\theta$ via $\varepsilon_c$ and substituting the result into (23) yields:

$$U(\theta) = 8 u_\theta \varepsilon_c^2 \cos^4(\theta_0/2). \qquad (24)$$

Here is the energy of stretching per one monomer unit due to the distortion of valence angles. The energy (24) multiplied by the number $\nu$ of monomers in unit volume should be equal to the crystal-like energy density (12a), i.e. $\nu U_\theta \approx W_c = 3/2 G_c \varepsilon_c^2$. This relation allows us to evaluate the modulus $G_c$ as follows:

$$G_c \approx u_\theta (3 + 4\cos\theta_0 + \cos 2\theta_0)/l^3 . \qquad (25)$$

Here we used the formula, $\cos^4(\theta_0/2) = 1/8(3 + 4\cos\theta_0 + \cos 2\theta_0)$, and the relation, $m \approx \rho l^3$ that have been employed before.

## 5. Crystallizing Polymers: Stress-Induced Crystallization

*5.1. Formation of the needle like (NL) crystals*

New effect, the stress-induced crystallization, occurs in stretching of crystallizing rubbery polymers. Our approach is based on the following hypothetical physical scenario. We assume that the special type of NL crystals emerge when some parts of macromolecules in well-stretched fibrils come randomly very close to each other. As soon as these fibrils are in a closed proximity of each other, the attractive forces cause these fibrils to suddenly coalesce in rigid NL crystals, whose shape is maintained by the emerged surface energy $\gamma$ acting on the crystal sidewall.

Consider the following deformed states for the same parts of macromolecules before and after formation of a NL crystal. The free energy density $W_e/\rho$ for amorphous fibrils before NL formation is described as: $W_e(\lambda_e) = 1/2 G_e \psi_e(\lambda_e)$ where we assume $\lambda_e \gg 1$. After formation of NL crystal, when the effect of surface energy emerges, the Gibbs's energy of a NL crystal is given by:

$$G_{\mathbf{c}} = \pi r_c^2 l_c W_e(\lambda_c) - 2\pi r_c l_c \gamma, \quad \lambda_c = l_c/l_0, \quad l_0 \approx \lambda_e l_d n_{\mathrm{P}}/n_d. \tag{26}$$

Here $\gamma$ is the surface energy coefficient, $r_c$ and $l_c$ are the radius and length of the crystal, $l_0$ is the length of stretched fibril before crystallization, and $n_{\mathrm{P}}$, an unknown parameter, is the number of monomer units in fibrils forming the crystal length,. Minimizing (26) with respect to $r_c$ and $l_c$ yields:

$$(\partial/\partial r_c) G_{\mathbf{c}} = 0: \quad r_c = \gamma/W_e(\lambda_c); \quad \lambda_c = \lambda_{em} \approx n_d^{1-s} \tag{27}$$

$$(\partial/\partial l_c) G_{\mathbf{c}} = 0: \quad \sigma(\lambda_c) = \gamma/r_c. \tag{28}$$

In obtaining (28) we used (27) and the definition of stress: $\sigma(\lambda_c) = \lambda_c \partial W_e/\partial \lambda_c$. We also used in (27), (28) the relation, $\lambda_c = \lambda_{em} \approx n_d^{1-s}$, because the macromolecular parts including in NL crystal are completely stretched.

On the other hand, the value of stress $\hat{\sigma}$ for a cylindrical rod stretched with the ratio $\lambda$, having a current radius $r$, and being under action of surface energy on the sidewall is:

$$\hat{\sigma} = \sigma(\lambda) - \gamma/r. \tag{29}$$

Formula (29) shows that the squeezing effect of surface tension produces the release of the true stretching stress. Comparing (28) and (29) yields the striking conclusion: *just formed NL crystal is completely released from the elongation force.* This reveals the compensatory mechanism of extensional strain release by surface energy and describes the occurrence of NL crystals as a spontaneous transition from oriented polymer to the NL crystals. As soon as the NL crystal is formed, it is loaded once again in the axial direction from the outside macromolecules, causing, however, almost no deformation in NL crystal. The important consequence of this analysis is that just formed NL crystals can be viewed in the following deformation history as *rigid fillers*. The *rigidity* of the NL crystal means that the additional (infinitesimal) stretching of formed NL crystals could only be caused by the fine motions of monomer units due to the distortion of valence angles.

It is evident that the asymptotic formulae (17), derived for the situation of highly stretched dynamic segments, are completely valid in the case of formation of NL crystal. Substituting the value of $G_e$ from (17) into the left-hand side of (25), reduces (25) to the form:

$$\frac{\Delta E_r}{l^3} \frac{\lambda_{em} \psi'_e(\lambda_{em})}{\psi_e(\lambda_{em})} = \frac{\gamma}{r_0}, \quad \text{or} \quad r_c = r_0 = \frac{\gamma l^3}{\Delta E_r} \cdot \frac{\psi_e(\lambda_{em})}{\lambda_{em} \psi'_e(\lambda_{em})}. \tag{30}$$

In case of validity (10a) and $\lambda_{em} \gg 1$, relation (30) is simplified to:

$$r_c = \gamma l^3 / \Delta E_r. \tag{31}$$

Here $r_c$ is the equilibrium radius of NL crystal.

The remarkable simple result (31) could be also readily derived when considering the Gibbs' energy function $G_c = \pi r_c^2 l_c \Delta E_r / l^3 - 2\pi r_c l_c \gamma$ for the single NL crystal. The first term in the right-hand side of this expression is the ultimate stretching energy equal to the energy of rotational barriers, overcome by monomer units confined in the NL crystal volume. The second term there is the surface energy of the single NL crystal. The value of the crystal radius $r_c$, which minimizes the Gibbs energy $G_c$, is found from (31), and respective minimum value of $G_c$ is represented as: $min\ G_c = -\pi l_c \gamma^2 l^3 / \Delta E_r$.

Although the equilibrium value of length $l_c$ of NL crystal cannot be quantitatively found using elementary approach, simple qualitative considerations show that

$$l_c \sim l_m = l\ n_d. \tag{32}$$

Indeed, the possible case, $l_c \ll l_m$, is improbable because of the high rigidity of dynamic segments. The opposite case, $l_c \gg l_m$, is improbable too because to form the NL single crystal of this high length several moderately extended dynamic segments should be almost perfectly align in the stretching direction.

Two conclusions could be drawn from our model of the NL crystal formation.
(i) The grow of a formed single NL crystal in the direction of extension is highly improbable. This is because at the instant of its formation, the crystal is relieved from the stretching. Although due to the action of the environmental polymer chains, the tension in the stretching direction occurs, it should be in general not enough to cause the grow of just formed NL crystal. The grow of a single NL crystal in the lateral direction is also forbidden due to the surface energy caused by the attractive intermolecular forces.
(ii) Unlike the thermally formed crystals obtained under cooling, the NL crystals formed under stretching do not create the "crystal phase" [14] but rather a mesophase. This is because polymer chains that are in amorphous state surround each NL crystal.

5.2. *The hybrid model for crystallizing rubbers*

In order to develop this hybrid model we need to specify only the free energy function and related expression for stress for energetic elasticity. Two other components of hybrid model, entropic and CL ones have been established in Section 3, with evaluation of their parameters given in Section 4. We remind that the typical approach of hybrid modeling is treating the phenomena in each of the three asymptotic regions of elasticity as independent.

In case of crystallizing rubbery polymers, the energetic component of elasticity might be easily considered as this type of elasticity for the filled system with NL crystals being the rigid filler. Then taking into account that the energetic elasticity happens in amorphous region of polymer deformation, and that in our simple modeling the NL crystals are oriented along the stretching direction, we can propose the modeling of energetic elasticity as:

$$W_e^c(\lambda_e,T) = W_e(\lambda_e,T)/(1-\varphi) = 1/2 G_e(T)\psi_e(\lambda_e) = 1/2 G_e(T)(2\lambda_e + \lambda_e^{-2} - 3)/(1-\varphi) \quad (33)$$

$$\sigma^c(\lambda_e) = \lambda_e \partial W_e^c / \partial \lambda_e \big|_{\varphi=const} = \sigma(\lambda_e)/(1-\varphi) = G_e(T)(\lambda_e - \lambda_e^{-2})/(1-\varphi). \quad (34)$$

Here $\varphi$ is the degree of NL crystalllinity, $W_e^c(\lambda_e,T)$ and $\sigma^c(\lambda_e)$ are free energy and stress, respectively, for crystallizing rubbers in "energetic region of deformations, $W_e(\lambda_e,T)$ and $\sigma(\lambda_e)$ are the corresponding variables for the amorphous polymers defined in (11a), (11b). It seems reasonable that in equilibrium case, $\varphi = \varphi(\lambda_e)$. Then the simplest phenomenological modeling of this dependence is:

$$\varphi(\lambda_e) = \chi \lambda_e / \lambda_{em}. \quad (35)$$

Here $\lambda_{em}$ is the ultimate energetic stretching ratio defined in (20) and $\chi$ (< 1) is the maximal possible degree of crystallinity treated here as an empirical factor. Substituting (35) into (34) yields:

$$\sigma^c(\lambda_e) = \frac{G_e(\lambda_e - \lambda_e^{-2})}{1 - \chi \lambda_e / \lambda_{em}}. \quad (36)$$

In the limit of exhausting energetic elasticity, when $\lambda_e \to \lambda_{em}$,

$$\sigma^c(\lambda_{em}) = \frac{G_e \lambda_{em}}{1-\chi}. \quad (37)$$

Using general formulae (4) and (6), it is now possible to establish the asymptotic formulae for hybrid elasticity for crystallizing rubbers, similar to those discussed in the Subsection 3.4. We consider here the case of strong energetic elasticity when $\lambda_e \gg 1$, with the relation (11d) being valid. In this case (10c) is valid too. Then the strain-stress relation obtained using (4) for the hybrid approach, is:

$$\lambda = \lambda_* \sqrt{\frac{\hat{\sigma}}{1+\hat{\sigma}}} \cdot \frac{\sigma}{\alpha + \chi \sigma / \lambda_{em}} \cdot \left(1 + \hat{\sigma} \frac{G_e}{3\alpha G_c}\right). \quad (38)$$

Here $\hat{\sigma}$ and $\alpha$ have been defined in (13a,b). Note that similarly to relations (13a,b), formula (38) is strictly valid when $\hat{\sigma} >> \alpha$, or equivalently, $\lambda_e >> 1$. However, one can start approximately using it when the variables in (38) have passed the crossover values defined in (14). Thus relation (38) describes the smooth transition from entropy to energetic elasticity for crystallizing rubber-like polymers. One should also mention that all the formulae for amorphous rubbers in Section 3 could be formally obtained from (38) and relation like that in the limit $\chi \to 0$. Note that due to (37), the formula (15c) that describes the transition from energetic to CL elasticity is also valid for the case of crystallizing rubbers in the vicinity $\lambda_e \approx \lambda_{em}$. Comparing the first two terms in the product in (38) with (13b) clearly shows that (38) describes the effect of reinforcement in stress-induced crystallization.

The result of a detailed analysis of the hybrid constitutive relation for crystallizing rubbers is sketched in Figure 3 where the strain-stress relation in crystallizing case is shown by the solid line and by dashed line for amorphous case. The effect of the stress-induced crystallization is well seen in Figure 3.

Analysis of another function, $\lambda_e(\lambda)$, based on the CE's of hybrid elasticity resulted in the dependence of the degree of crystallinity on the total stretching ratio, $\varphi(\lambda) = \chi \lambda_e(\lambda) / \lambda_{em}$, sketched in Figure 4. This plot obtained using modeling (35), demonstrates the occurrence of a quasi-threshold for onset of the stress-induced crystallization, because the "energetic" stretching ratio $\lambda_e$ is noticeable only when the entropic elasticity is well developed and/or almost exhausted.

Both the qualitative predictions in Figures 3 and 4 seem realistic as compared to the data in [1](ch.1).

## 6. Conclusions

The present paper develops a model of hybrid entropy/energetic elasticity on the example of simply stretching. The model employs a version of bid-spring model with deformable bids treated as dynamic segments. The extension of bids is attributed to the energetic elasticity dominated inside of dynamic segments, and is caused by collective motions of monomer units overcoming rotational barriers. Even when the energetic component in the large elasticity of cross-linked rubbers seems to be significant, more pronounced effects could be seen in the deformation of polymers in the rubbery state just above $T_g$. Though the deformation of polymers in this region is in non-equilibrium, the higher viscosities and relaxation times at these low temperatures could considerably elevate intensity of equilibrium part of stress.

Three different types of elasticity have been taken into account in formulation of the hybrid CE: (i) entropy type with finitely extendable chains, (ii) energetic type with finite elasticity known from description of low molecular weight hard materials, and (iii) crystal-like very tough elasticity known for minerals and metals.

The hybrid CE equations obtained here for simple stretching, is easily extended to the 3D tensor form. This 3D formulation will be published elsewhere.

Although only equilibrium type of modeling has been considered in this paper, it is needless to say that such non-equilibrium effects as relaxations (for cross-linked polymers) and both relaxation and flow effects (for not cross-linked polymers above $T_g$) could highly affect the mechanical and optical behavior of polymers in rubbery state.

**Acknowledgement**

Many thanks go to Professor M. Cakmak who generously discussed his mostly unpublished experimental results with the author.

**Figure Captions**

Fig.1. A – common bid-spring model, B - bid-spring model with extensible bids.

Fig. 2. Schematic plots of $\hat{\lambda}(\hat{\sigma};\alpha)$ for various values of $\alpha$, $\alpha_1 < \alpha_2 < \alpha_3$.

Fig.3. Schematic plots of strain-stress dependences $\lambda(\sigma)$ for crystallizing (solid line) and amorphous (dashed line) rubber-like materials.

Fig.4. Sketch of dependence $\varphi(\lambda)$ of degree of crystallinity on the stretching ratio, qualitatively predicted by the hybrid approach.

Fig.1

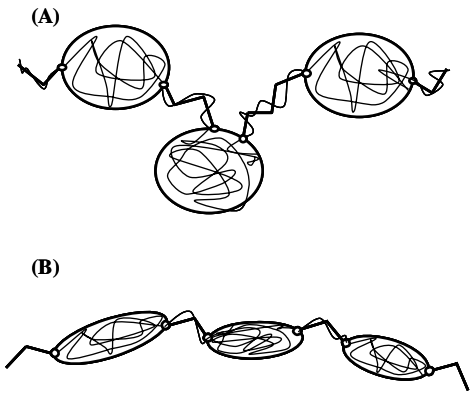

Fig.3

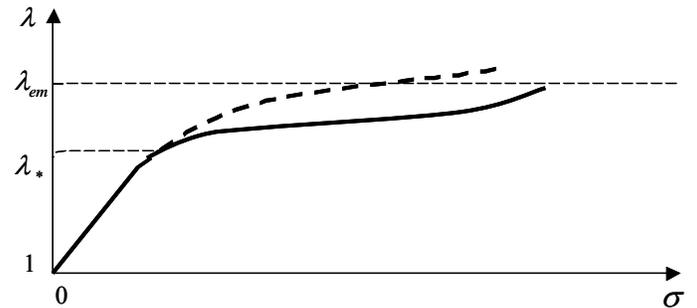

Fig.2

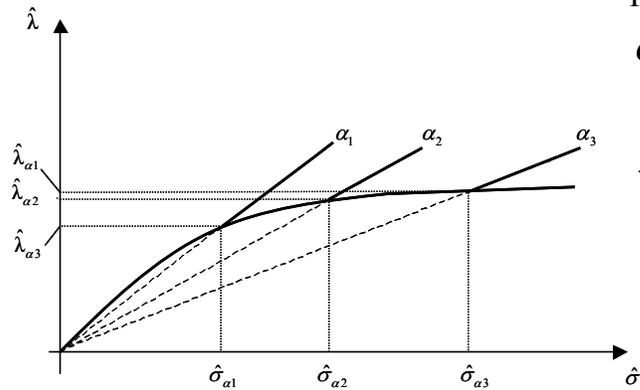

Fig.4

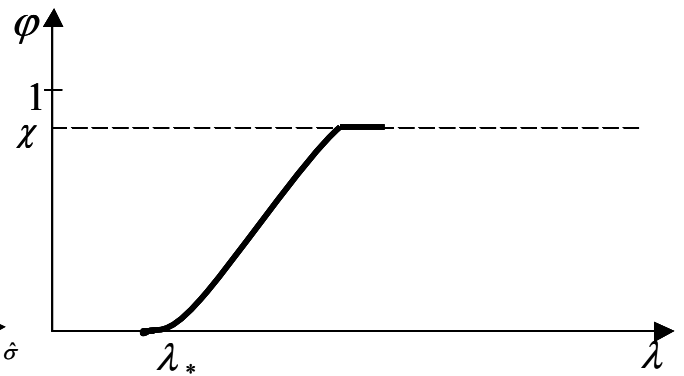